\documentclass{emulateapj}

\newcommand{\til}{$\sim$}
\newcommand{\ergsqcmsec}{\thinspace\hbox{$\hbox{erg}\thinspace\hbox{cm}^{-2}
                \thinspace\hbox{s}^{-1}$}}

\def\spose#1{\hbox to 0pt{#1\hss}}
\def\simlt{\mathrel{\spose{\lower 3pt\hbox{$\mathchar"218$}}
     \raise 2.0pt\hbox{$\mathchar"13C$}}}
\def\simgt{\mathrel{\spose{\lower 3pt\hbox{$\mathchar"218$}}
     \raise 2.0pt\hbox{$\mathchar"13E$}}}

\def\today{\ifcase\month\or
January\or February\or March\or April\or May\or June\or
July\or August\or September\or October\or November\or December\fi
\space\number\day, \number\year}

\slugcomment{Accepted for publication by the Publications of the Astronomical Society of the Pacific,  December 16, 2004}

\shorttitle{SDSS J2100}
\shortauthors{Tramposch et al.}

\begin{document}


\title{SDSS~J210014.12+004446.0: A New Dwarf Nova with Quiescent Superhumps?$^1$}


\altaffiltext{1}{Based on observations obtained with the Sloan Digital Sky Survey and with the Apache Point Observatory (APO) 3.5m telescope, which are owned and operated by the Astrophysical Research Consortium (ARC)}
\author{Jonica Tramposch\altaffilmark{2}, Lee Homer\altaffilmark{2}, Paula Szkody\altaffilmark{2}, Arne Henden\altaffilmark{3,4}, Nicole
  M. Silvestri\altaffilmark{2}, Kris Yirak\altaffilmark{2}, Oliver J. Fraser\altaffilmark{2}, J. Brinkmann\altaffilmark{5}} 
\altaffiltext{2}{Department of Astronomy, University of Washington, Box 351580, Seattle, WA 98195}
\email{jonica@u.washington.edu; homer@astro.washington.edu; szkody@astro.washington.edu; aah@nofs.navy.mil;
  nms@astro.washington.edu; yi\nolinebreak rak@pas.rochester.edu;
  fraser@astro.washington.edu; jb@apo.nmsu.edu}
\altaffiltext{3}{US Naval Observatory, Flagstaff Station, P.O. Box 1149, Flagstaff, AZ 86002-1149}
\altaffiltext{4}{Universities Space Research Association}
\altaffiltext{5}{Apache Point Observatory, 2001 Apache Point Road, P.O. Box 59, Sunspot, NM 88349-0059}




\begin{abstract}

We report follow-up observations of the Sloan Digital Sky Survey Cataclysmic Variable SDSS~J210014.12+004446.0 (hereafter SDSS J2100).  We
obtained photometry and spectroscopy in both
outburst and quiescent states, providing the first quiescent spectrum of this source.  In both states, non-sinusoidal photometric modulations are apparent, suggestive of
superhumps, placing SDSS~J2100 in the SU UMa subclass of dwarf novae.  However, the periods during outburst and quiescence differ
significantly, being $2.099\pm0.002$~hr and $1.96\pm0.02$~hr respectively.  Our phase-resolved spectroscopy during outburst yielded an estimate
of \til2 hr for the orbital period, consistent with the photometry. The presence of the
shorter period modulation at quiescence is unusual, but not unique.  Another atypical feature is the relative weakness of the Balmer emission
lines in quiescence.  Overall, we find a close similarity between SDSS~J2100 and the well-studied superhump cataclysmic Variable V503 Cygni.  By analogy,
we suggest that the quiescent modulation is due to a tilted accretion disk -- producing negative superhumps -- and the modulation in outburst is due to
positive superhumps from the precession of an elliptical disk. 
\vspace*{7mm}
\end{abstract}

\keywords {cataclysmic variables --- stars: individual (SDSS J210014.12+004446.0) --- accretion, accretion disks}

\section{Introduction}

Cataclysmic variables (CVs) are binary systems in which a white dwarf accretes material from a donor (secondary) star, due to Roche-lobe
overflow \citep[see review by][]{warn95}.  In non-magnetic CVs, the accreting material from the donor forms a disk around the white dwarf.  Superhumps -- periodic modulations of the optical flux -- are attributed to this disk being asymmetric and/or tilted out of the orbital plane \citep{warn95}.

There are several types of superhumps \citep[see][appendix]{patt02}.  Relevant to this paper are the orbital hump, common, permanent, and
quiescent superhumps.  The orbital hump is a signal in quiescence at the binary orbital frequency; the common superhump -- what is
usually assumed when referring to superhumps -- is a large amplitude signal whose period is a few percent longer than the orbital, decays in a
matter of weeks, and is displayed in all SU UMa stars in outburst; the permanent superhump lasts for months (or longer) and is not associated
with eruption; quiescent superhumps are rare, occur during phases of low luminosity, and have no apparent connection to other types of
superhumps \citep{patt02}.  In terms of observational terminology, positive and negative refer to superhumps having periods greater or smaller
than that of the binary's orbital period.  Positive superhumps are currently explained by the prograde precession of an elliptical accretion
disk, relative to the orbit \citep{whit88,osak89,lubo91}. It is proposed that negative superhumps arise when the disk also extends out of the orbital plane and the line of nodes undergoes retrograde precession due to the
gravitational torque of the donor star \citep{harv95,patt97,patt99,wood00,murr02}.  Positive and negative superhumps have been observed to coexist \citep[see][]{harv95}, though they may be present separately.
\begin{figure}[!tb]
\resizebox{.46\textwidth}{!}{\rotatebox{0}{\plotone{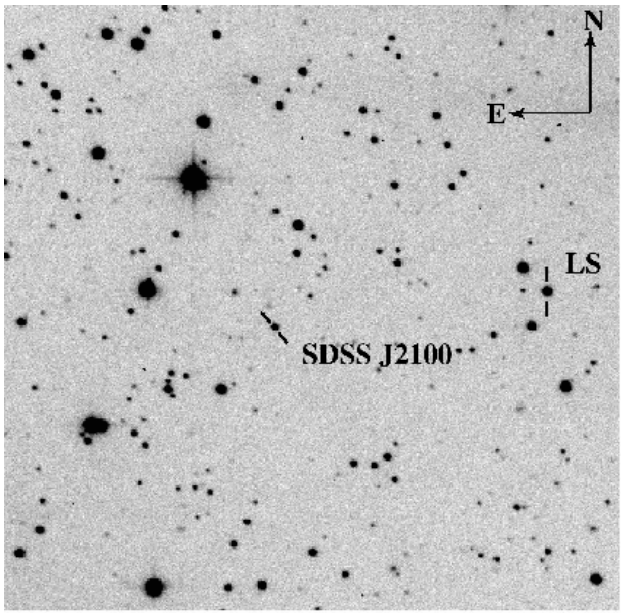}}}
\caption {Finding chart for SDSS~J2100.  The image is $8.5'\times 8.5'$ and is taken in the Harris $V$ filter.  The local standard star used for the differential photometry from MRO is marked as LS. \label{fig:find}}
\end{figure}

The Sloan Digital Sky Survey (SDSS) has completed over three years of a photometric imaging and spectroscopic surveying  \citep{abaz03,abaz04,fuku96,gunn98,hogg01,lupt99,lupt01,pier03,smit02,stou02,york00}.
Although the prime aim of the spectroscopy is to measure redshifts of over one million galaxies and quasars, the color overlap between the quasars and
CVs has resulted in numerous new identifications from the SDSS spectra. To date, 88 new CVs have been discovered
\citep[see][]{szko02b,szko03,szko04}, including the system SDSS~J210014.12+004446.0 (for brevity, hereafter SDSS~J2100).  The SDSS spectrum of SDSS~J2100 \citep[see][]{szko04} exhibits the narrow Balmer emission and blue continuum typical of dwarf nova (DN) in outburst.
Moreover, earlier Sloan photometry showed a magnitude of $g=18.8$ but the SDSS spectrum is two magnitudes brighter, with a flux level equivalent to $V=16.7$.  We therefore obtained follow-up time-resolved photometry of SDSS~J2100
at two different epochs, as well as a spectrum in quiescence to confirm its classification as a DN and to further delineate its characteristics.

\section{Observations and Results}

\begin{deluxetable*}{llllcl}
\tablewidth{0pt}
\tablecaption{Observation Summary\label{tab:obslog}
}
\tablehead{\colhead{UT Date} & \colhead{Julian Date} & \colhead{Obs} & \colhead{UT Time} &\colhead{$\overline{V}$}&
\colhead{Comments}}
\startdata
2003 Jul 26 & 2452846 & MRO &  05:17--11:23 &16.3& $V$ photometry\\
2003 Jul 27 & 2452847 & MRO & 04:51--11:24 &16.5&  $V$ photometry\\
2003 Oct 19 & 2452931 & NOFS & 01:32--06:53&18.3& open filter photometry\\
2003 Dec 30 & 2453003 & APO/DIS & 01:14--02:32& 18.7 & spectroscopy\\
2004 Oct 13 & 2453291 & APO/DIS & 01:43--04:24& 17 & spectroscopy\\
\enddata
\end{deluxetable*}

\begin{figure*}[!htb]
\begin{center}
\resizebox{.60\textwidth}{!}{\rotatebox{-90}{\plotone{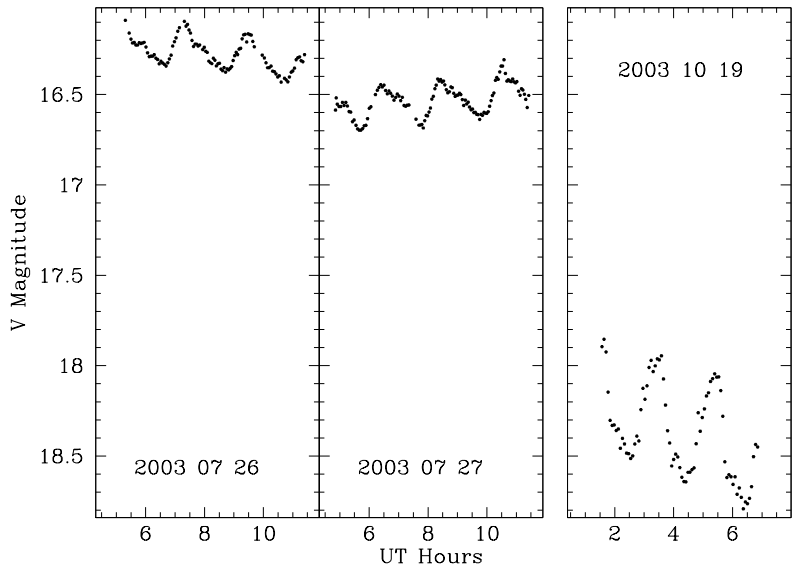}}}
\caption{Differential photometry of SDSS~J2100 in outburst from MRO ($V$ band) and in quiescence NOFS (unfiltered, but with $V$ band zeropoint). \label{fig:lcs}}
\end{center}
\end{figure*}

SDSS~J2100 was observed with the 0.76m telescope at Manastash Ridge Observatory (MRO) on 2003 July 26 and 27.  During these two nights, 226 exposures of 150--180s (giving
a typical time resolution \til200s) were
taken in the Harris $V$ filter.
Aperture photometry was performed on the reduced images within
IRAF\footnote{IRAF (Image Reduction and Analysis Facility) is distributed by the National Optical Astronomy Observatories, which are operated
  by the Association of Universities for Research in Astronomy (AURA) Inc., under cooperative agreement with the National Science Foundation}. Differential magnitudes were then calculated relative to a brighter local standard star in the field (see
Figure~\ref{fig:find}).  Photometry was obtained for three other stars of comparable brightness to verify stable
photometric conditions.  All four stars possess calibrated Johnson $V$ magnitudes from nights of all-sky photometry, including Landolt standards,
and were used to provide the $V$ band zero point.  A summary of the observations is given in Table~\ref{tab:obslog}.

Photometric observations of SDSS~J2100 were also undertaken with the US Naval Observatory Flagstaff Station (NOFS) 1m telescope on 2003 October
19 (see Table~\ref{tab:obslog}).  Longer exposures of 210s were taken, giving 250s time resolution including dead-time.  Differential aperture
photometry with no filter, relative to an ensemble of 14 local calibrated standard stars, was used to obtain an approximate $V$ band zeropoint.  The lightcurves from both sets of observations are plotted in Figure~\ref{fig:lcs}.

\begin{figure}[!tb]
\resizebox{.46\textwidth}{!}{\rotatebox{0}{\plotone{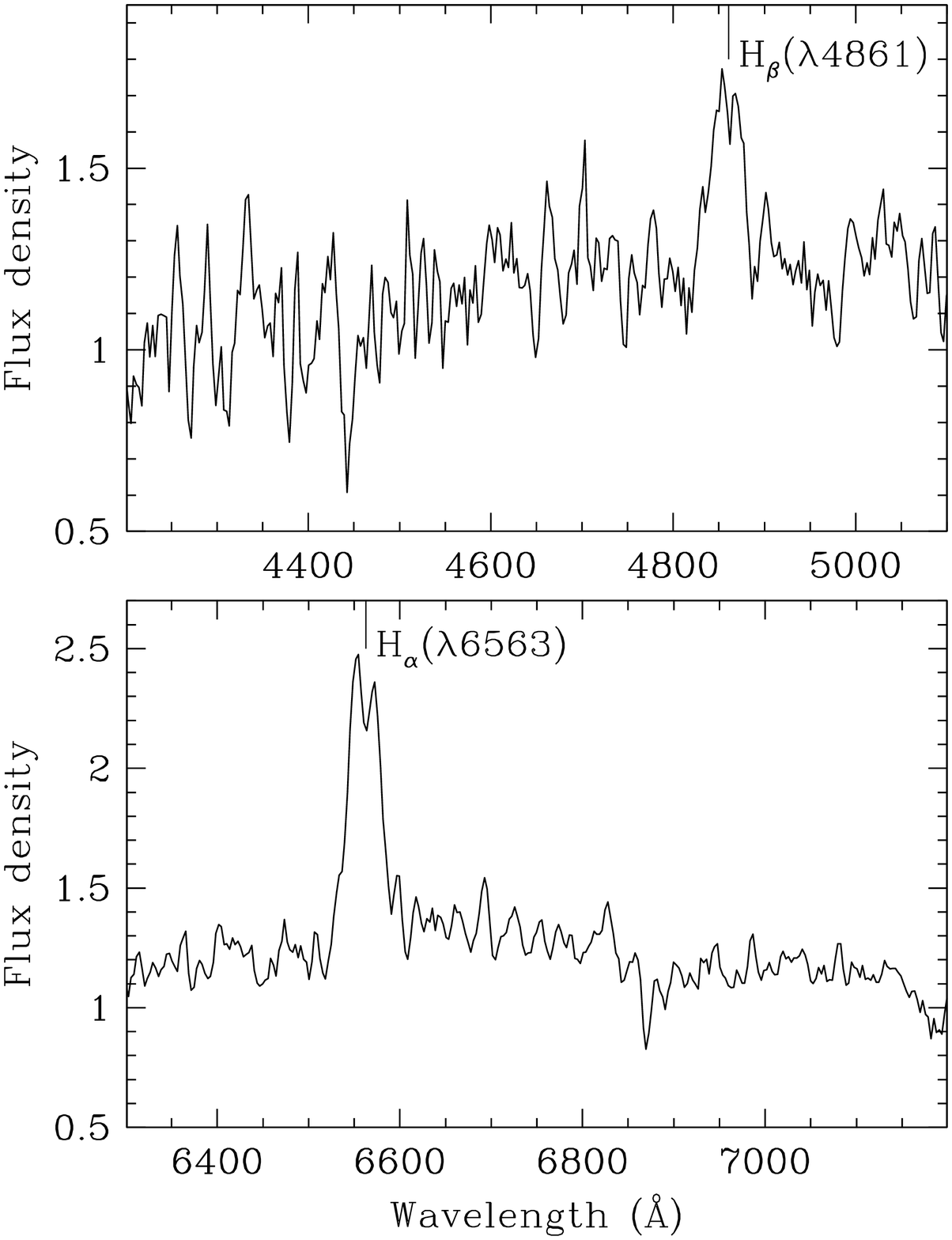}}}
\caption{A low-resolution spectrum of SDSS~J2100 in the faint state.  The data have been binned to 3\AA\, then box car smoothed by three for
  clarity. Double-peaked Balmer lines typical of DN in quiescence are present, although the
  lines are weaker than for most such systems. The flux density is in units of $10^{-16}$\ergsqcmsec. \label{fig:spec}}
\end{figure}

On 2003 December 30 and 2004 October 13, we obtained sets of spectra using the Double-Imaging Spectrograph (DIS) on the 3.5m telescope at Apache Point
Observatory (APO; see Table~\ref{tab:obslog}).  The exposure times ranged from 900 s to 1200 s.  Given the faintness of the target in 2003 December, the four spectra were co-added to yield sufficient signal-to-noise to
identify and measure the emission lines; hence, no time/phase resolution is available. In Figure~\ref{fig:spec} we present the resulting low-resolution
(binned to 3\AA) spectrum of SDSS~J2100 in quiescence.  The only prominent lines are those of H$\alpha$ and
H$\beta$, both of which appear double-peaked; we measure equivalent widths of $37\pm2$\AA\ and $14\pm1$\AA\ respectively.  In 2004 October,
the lines were still weak, with H$\alpha$ being 15.9\AA\ and H$\beta$ at 5.3\AA, although the system was almost 2 magnitudes brighter.   The radial
velocity curves (from 13 spectra spanning 2.5 hr) exhibit much scatter, only enabling us to estimate the orbital period as \til2 hr.  We likely caught SDSS~J2100 on decline from outburst maximum at
this time -- both its magnitude and line strengths are intermediate between the 2003 December observation and the original SDSS spectroscopy \citep[see][]{szko04}.

From Figure~\ref{fig:lcs} and Table~\ref{tab:obslog}, it is clear that SDSS~J2100 is highly variable on timescales from hours to months.  The
average magnitudes differ by \til2 between bright state (2003 July and 2004 October) and
faint state (2003 October and 2003 December)
observations.  However, in both photometric observations, large amplitude variations on \til2 hr timescale are apparent, as are significant longer term
trends.  

\begin{figure}[!tb]
\resizebox{.46\textwidth}{!}{\rotatebox{0}{\plotone{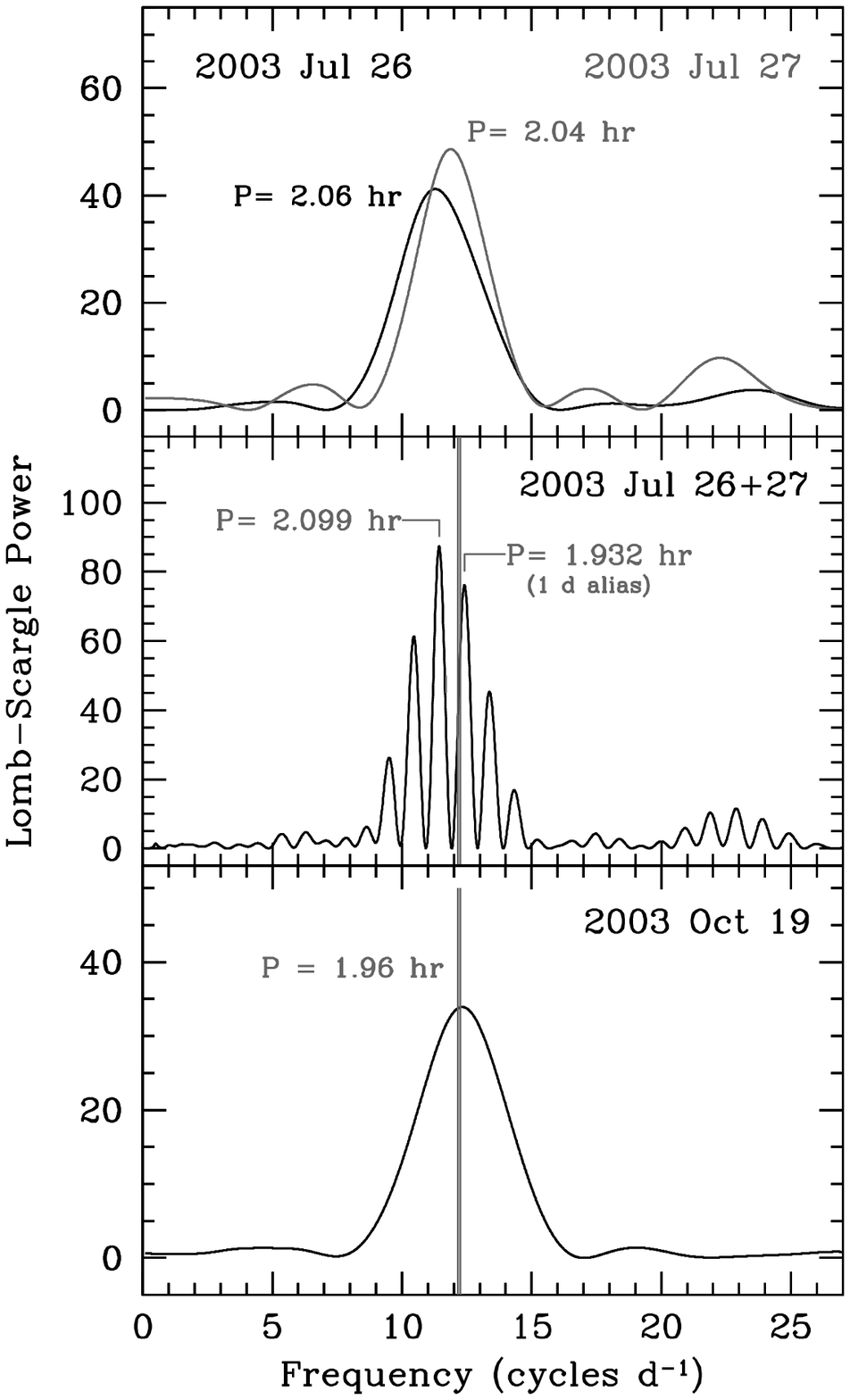}}}
\caption{Lomb-Scargle Periodograms of the de-trended lightcurves of SDSS~J2100 (from top to bottom): 2003 July 26 (black) and 2003 July 27 (grey),
  2003 July 26 \& 27 combined, 2003 October 19. The refined periods -- from fitting of a sinusoid + first harmonic model to the
  lightcurves -- are indicated. \label{fig:lsps}}
\end{figure}

\begin{figure}[!tb]
\resizebox{.46\textwidth}{!}{\rotatebox{0}{\plotone{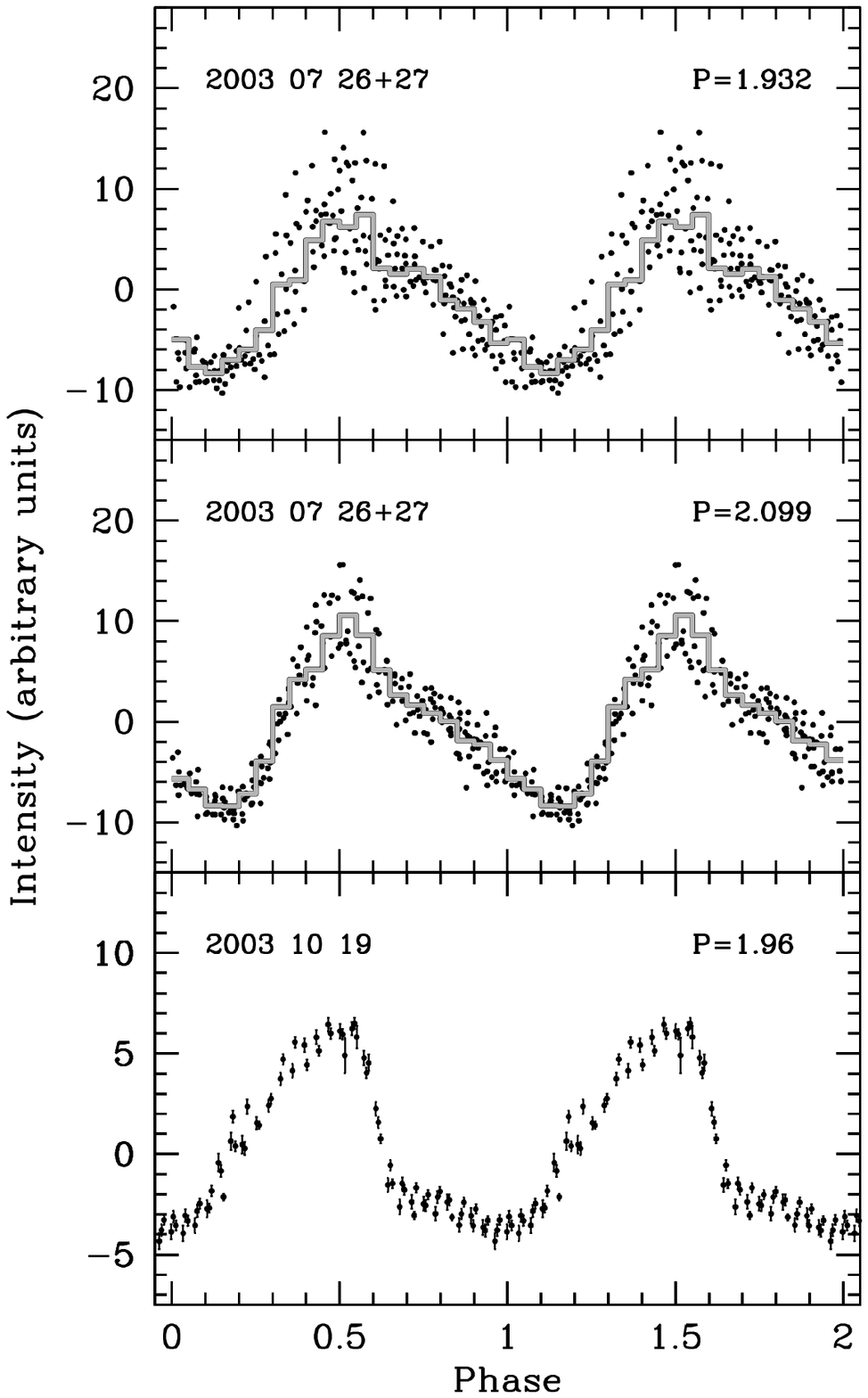}}}
\caption{Lightcurves of SDSS~J2100 phase-folded on measured periods.  Before folding, each of the light curves was de-trended.  From top to
  bottom: combined July dataset folded on the shorter period alias, on the best fit period, and the October lightcurve folded on its best fit
  period (note this panel is plotted with half the intensity range of the other two).  The phase-binned lightcurves are over-plotted on the July data.  For explanation of intensity scale see text.\label{fig:fds}}
\end{figure}

To search for periodicities, we first removed the nightly long-term trends visible in Figure~\ref{fig:lcs}.  For the two July datasets combined, we
modeled the trend by \til3~d sinusoid.  For the October
curve we used a simple linear fit.  Using these de-trended data, we calculated Lomb-Scargle Periodograms \citep{scar82}
for each night and for the July dataset combined (see Fig.~\ref{fig:lsps}).  Given the limited data spans relative to the \til2 hr modulation
and non-sinusoidal morphology, we then used least squares
fitting of a sinusoid plus the first harmonic model to refine the period determinations;  we 
included a sinusoid on the order of days and a linear component as free parameters to fit the longer term trends in each of the {\em original} July and October datasets respectively.  
From the combined July dataset we find a best fit period of $2.099\pm0.002$ hr. The one day alias at $1.932\pm0.002$ also fits the data, but
with a $\chi^2$ almost twice as large (also see increased scatter about mean phase folded curve in Fig.~\ref{fig:fds}), we discount this option.  In October, the best fit model
yields a period of  $1.96\pm0.02$ hr, significantly different from the best fit in outburst.  In all cases, the errors are derived from the least-squares fitting, but were
also double-checked in the frequency domain.  We generated $\sim$20,000 fake light curves, consisting of the de-trended, signal-subtracted magnitude values
randomized over the time points, plus a simple sinusoid (with randomized phase) at the fundamental frequency.  The width of the resulting peak
frequency distribution from the respective Lomb-Scargle Periodograms agreed with the light curve fits.  We also note that aside from the 27 July data set, the period values were insensitive (within the quoted errors) to different choices of reasonable de-trend models.  For the 27 July data set we conservatively quote an appropriately larger uncertainty.

Could this shorter 1.96~hr period be present in the July dataset?  In frequency space, it is sufficiently
close to the 1.932~hr alias that confusion could occur.  However, considering the time domain, if we require the hump profile to remain approximately the same from one night to the next, the
full amplitude of any 1.96~hr signal must be $\simlt0.06$ mag.  

From Figure~\ref{fig:lcs}, it is clear that the \til2~hr
modulations are non-sinusoidal and their morphology varies between July and October.  Even more striking is the difference in the amplitudes --
they vary from 0.22~mag (peak to peak) in the bright state to 0.63~mag in the faint.  To further examine the changes and periods, we folded the de-trended lightcurves
on their candidate periods (additionally  phase binning the July data) and converted the magnitudes to normalized intensity units, such
that an intensity of 100 corresponds to the average $V$
magnitude of 16.3 on 2003 July 26  (see Fig.~\ref{fig:fds}).  Here we can clearly see the change in morphology from steeper rise to steeper
decline between the two observations.  Lastly, it is interesting to note that in terms
of intensity, the humps are twice as large in the bright state compared to the faint, with full amplitudes 20.2 (phase binned) versus 9.7. The period measurements and amplitudes are summarized in Table~\ref{tab:res}.

\begin{deluxetable*}{llclcl}
\tablewidth{0pt}
\tablecaption{Period Determinations\label{tab:res} }
\tablehead{\colhead{Dataset} &\colhead{Period (hr)} & \colhead{Mean Intensity\tablenotemark{a}} & \multicolumn{2}{c}{Full Amplitude} &
\colhead{Comments}\\
&&&intensity& mags&}
\startdata
2003 Jul 26 & $2.09\pm0.04$ & 100 & $20.2\pm1.2$&$0.22 \pm0.01 $ & Positive superhump\\
2003 Jul 27 & $2.04\pm0.01$ & 83 &$20.0\pm1.0$&$0.27 \pm0.01 $ & Positive superhump\\
2003 Jul 26--27 & $2.099\pm0.002$& 89 &$19.7\pm0.7$& $0.24 \pm0.01 $& Best fit ($\chi^2_{\nu}=2.4$)\\
                & $1.932\pm0.002$& 89 &$15.7\pm0.7$& $0.19\pm0.01$& 1-day alias ($\chi^2_{\nu}=4.3$)\\
               & 1.96 (imposed) & 89 & $\simlt5$ &$\simlt0.06$ & Amp. limits \\
2003 Oct 19 & $1.96\pm0.02$& 16 &$9.7\pm0.2$& $0.63\pm0.02$& Negative superhump?\\

\enddata
\tablenotetext{a}{The intensity scale is normalized such that a value of 100 corresponds to the mean magnitude ($V=16.3$) of the first dataset.}
\end{deluxetable*}

 \section{Discussion}
Considering the average magnitudes of SDSS~J2100 at the various epochs, the 2 mag difference between observations is suggestive of the range seen in short-period DN, although we
note that it is 
smaller than most \citep[see][]{warn95}.  The
original SDSS photometry,
Flagstaff photometry, and the 2003 December APO spectrum were likely taken in (or in approach to) quiescence, whereas it was observed near outburst during the
original SDSS spectroscopy, at MRO, and at APO in 2004 October.  The fact that we have observed SDSS~J2100 three-times at $V\sim16.7$ suggests that we caught it close to
maximum.  Moreover, the significant non-sinusoidal humps in our bright state photometry have the classic appearance of superhumps, as seen in SU
Ursae Majoris type DN during superoutburst.  The observed period of 2.10~hr for SDSS~J2100 at outburst and the 1.96~hr period in quiescence are
sufficiently different (7$\sigma$) that they cannot be the same period; i.e., at least one of them must be a superhump.  Similar quiescent periodicities shorter than those in outburst have been seen in the SU UMa stars: V503~Cygni and IR Gem
\citep{harv95,fuxx04}, where they were attributed to negative superhumps. 

It is instructive to compare the temporal properties of SDSS~J2100 to those of the better studied SU UMa DN, V503~Cygni.  V503~Cygni varies over
four magnitudes from quiescence to maximum and has an orbital period of $1.865\pm 0.007$ hr, as determined by a radial velocity study \citep{harv95}.  From
extensive photometric observations, Harvey et al. identified positive superhumps during superoutburst with a period of $1.945 \pm0.002$ hr
and peak-to-peak amplitude of 0.13 mag.  More
interestingly, V503~Cygni exhibited periodic modulations in both outburst {\em and} quiescence with a period of $1.8167\pm0.0002$ hr, which was identified
as a negative superhump. The amplitude of this negative superhump signal varied from 0.034 mag in outburst to 0.5 mag in quiescence; in intensity units, however, the amplitudes of the negative superhumps were found to be constant within 30\%.  
With a positive superhump period of 2.10~hr, let us consider the two options for the origin of the 1.96~hr signal in SDSS J2100: modulation on the orbital period, or a negative superhump like in
V503~Cygni.  If it is an orbital modulation, it is most readily explained as the varying light due to a hotspot.  Observing this
would require a fairly high inclination ($\sim70^\circ$), which is consistent with the pronounced central absorption of the emission lines in our outburst spectroscopy.  On the other hand, such an identification would yield a superhump excess of $\epsilon_+=(P_{\rm sh}-P_{\rm
  orb})/P_{\rm orb} = 0.07$, greater than any of the \til2~hr superhumps listed by \citet{patt99}, which are around 0.04. Instead,
if the 1.96~hr signal is a negative superhump, the period difference is very close to other \til2~hr two-mode superhump variables, V503~Cygni (with 6.5\%) and
V1974 Cyg \citep[7.3\%][]{rett97}.  We can also use the two superhump excesses, taking $\epsilon_-/\epsilon_+=-0.56\pm0.06$ from \citet{rett02},
to estimate the orbital period as $2.0\pm0.1$ hr (where $\epsilon_-=-0.025$ hr and $\epsilon_+=0.044$ hr).

\begin{deluxetable}{llll}
\tablewidth{0pt}
\tablecaption{Dwarf Novae Equivalent Widths in Quiescence\label{tab:EWs} }
\tablehead{\colhead{Source} &\colhead{H$\alpha$ (\AA)} & \colhead{H$\beta$ (\AA)} &
\colhead{Reference} }
\startdata
V503~Cygni & 30--40 & 13--23 & \citet{harv95}\\
SDSS~J2100 & 37 & 14 & This paper\hspace*{0.8cm}\\
SX LMi &  83 &  52--73 & \citet{wagn98}\\
\enddata
\end{deluxetable}

Since the negative superhump in V503~Cygni was observed during outburst and quiescence, it qualifies as a permanent superhump.  Such a
distinction cannot yet be made for the SDSS J2100 negative superhump -- it could either be a purely quiescent or a permanent superhump.  The
positive superhumps for both systems suggest common superhumps, given their large amplitudes and, in the case of V503~Cygni, a decay time on the order of weeks.

If we take the 1.96 hr period to be a negative superhump, the positive superhump has an intensity amplitude 2 times greater than that of the quiescent negative superhump.  For V503~Cygni, we
estimate this ratio to be 2--5, which is comparable.  However, there is one notable difference between the superhumps in the two systems.  At outburst, removal of the fundamental plus the first harmonic of the
2.1~hr signal from SDSS J2100 does not reveal any signal near 1.96~hr, with a limiting amplitude roughly half of that in quiescence (in intensity units).
Such a negative superhump was clearly discernible in the V503~Cygni outburst data after similar subtraction of the positive superhump. Perhaps the lack of this in SDSS J2100 can be explained by our small data span.  It is
conceivable, however, that in July the negative superhump had a shape/amplitude more easily hidden beneath the large positive signal -- we 
note that the shape of the negative superhump in V503~Cygni is far from stable, appearing to evolve over time.  Such morphological variations may relate to the change
in disk size between quiescence and (super)outbursts, or could be a cyclic feature, recurring on the precession period of the tilted disk, as
is seen in the dips of the low-mass X-ray binary with positive and negative superhumps, X1916-015/V1405 \citep{home01a,rett02}. 


There is one further similarity between SDSS~J2100 and V503~Cygni: the emission lines seen in
Figure~\ref{fig:spec} appear fairly weak compared to the continuum level.  Indeed, the equivalent widths are lower than for most SU UMa systems
in quiescence.

In Table~\ref{tab:EWs} we summarize the  results for H$\alpha$ and H$\beta$ for SDSS~J2100, V503~Cygni and a typical SU UMa DN, SX LMi. Our
results for SDSS~J2100 lie well within the range measured for V503~Cygni.  They are both less than half the equivalent widths for SX LMi.

\section{Conclusions}
We have presented evidence to support the classification of SDSS~J2100 as a new dwarf nova; more notably, one that exhibits negative superhumps in quiescence.  In
outburst photometry, we detect positive superhumps, placing
it among the SU UMa group; this is as expected, since its period places it (just) below the period gap \citep{diaz97}.  Both photometrically and spectroscopically,  SDSS~J2100 appears to
be 
an analog of the 1.9 hr system V503~Cygni, showing similar large amplitude variability and relatively weak Balmer lines in quiescence.  

Additional spectroscopy and photometry are needed to further probe this unusual system.  First, more extensive photometry should be done to confirm and
further constrain the positive and negative superhump periods.  Especially useful would be the detection of the negative superhump during
outburst (this will complete the analogy with V503 Cyg).  Second, more extensive, higher signal-to-noise phase-resolved spectroscopy will enable the system's orbital period to be determined. 

\acknowledgments

Funding for the creation and distribution of the SDSS Archive has been provided by the Alfred P. Sloan Foundation, the Participating Institutions,
the National Aeronautics and Space Administration, the National Science 
Foundation, the U.S. Department of Energy, the Japanese
Monbukagakusho, and the Max Planck Society. The SDSS Web site is 
http://www.sdss.org/.

The SDSS is managed by the Astrophysical Research Consortium (ARC) for the 
Participating Institutions. The Participating Institutions are The
University of Chicago, Fermilab, the Institute for Advanced Study, the 
Japan Participation Group, The Johns Hopkins University, the Korean Scientist Group, Los Alamos
National Laboratory, the Max-Planck-Institute for Astronomy (MPIA), the 
Max-Planck-Institute for Astrophysics (MPA), New Mexico State
University, University of Pittsburgh, Princeton University, the US 
Naval Observatory, and the 
University of Washington.

\end{document}